\documentclass[fleqn,10pt]{wlscirep}
\usepackage[utf8]{inputenc}
\usepackage[T1]{fontenc}
\usepackage{amsmath}
\title{The toroidal field surfaces in the standard poloidal-toroidal representation of magnetic field}

\author[1]{Sibaek Yi}
\author[1,2,*]{G. S. Choe}

\affil[1]{School of Space Research, Kyung Hee University, Yongin 17104, Korea}

\affil[2]{Department of Astronomy \& Space Science, Kyung Hee University, Yongin 17104, Korea}

\affil[*]{gchoe@khu.ac.kr}

\begin{abstract}
The representation of magnetic field as a sum of a toroidal field and a poloidal field has not rarely been used in astrophysics, 
particularly in relation to stellar and planetary magnetism. 
In this representation, each toroidal field line lies entirely in a surface, 
which is named a toroidal field surface. 
The poloidal field is represented by the curl of another toroidal field and it threads a stack of toroidal field surfaces. 
If the toroidal field surfaces are either spheres or planes, the poloidal-toroidal (PT) representation is known to have a special property 
that the curl of a poloidal field is again a toroidal field . 
We name a PT representation with this property a standard PT representation while one without the property is called a generalized PT representation. 
In this paper, we have addressed the question whether there are other toroidal field surfaces allowing a standard PT representation than spheres and planes. 
We have proved that in a three dimensional Euclidean space, there can be no standard toroidal field surfaces other than spheres and 
planes, which render the curl of 
a poloidal field to be a toroidal field. 

\end{abstract}

\begin{document}

\flushbottom
\maketitle
%
%
\thispagestyle{empty}


\section*{Introduction}

Although a magnetic field ${\bf B}$ has three components, they are not independent of each other 
due to the constraint $\nabla \cdot {\bf B} = 0$, which allows us to describe the magnetic field 
by two scalar fields only. 
Among such descriptions, the most well-known one has the form
\begin{equation} \label{eq:euler}
{\bf B} = f(\alpha, \beta) \nabla \alpha \times \nabla \beta \, , 
\end{equation}
in which two scalar fields $\alpha$ and $\beta$ are called Euler potentials or Clebsch variables
\cite{Stern70, Stern76, Dhaeseleer91} and $f$ is an arbitrary function of two variables $\alpha$ and $\beta$.  
As can be seen in Eq. (\ref{eq:euler}), a field line is defined as the intersection of 
a constant $\alpha$ surface and a constant $\beta$ surface. The Euler potentials, however, 
may not be single-valued for certain global fields, which limits their use for general magnetic field description \cite{Stern70}.

A more general two scalar description of magnetic field is the poloidal-toroidal respresentation 
(hereafter PT representation)  
\cite{Elsasser46, Lust54, Chandrasekhar57, Chandrasekhar61, Backus58, Stern76, Backus86, Low06, Low15, Berger18}, 
also called the Mie representation \cite{Backus86} or the Chandrasekhar-Kendall representation \cite{Low06}.
In this description, a magnetic field is decomposed into 
two divergence-free (solenoidal) fields, a poloidal field ${\bf B}_P$ and a toroidal field ${\bf B}_T$, i.e., 
\begin{equation} \label{eq:bpt}
{\bf B} = {\bf B}_P + {\bf B}_T \, ,
\end{equation}
in which
\begin{equation}
\label{eq:bp}
{\bf B}_P = \nabla \times ( \nabla \xi \times \nabla \Phi )
= \nabla \times ( \nabla \times \xi \nabla \Phi ) = - \nabla \times ( \nabla \times \Phi \nabla \xi )\, ,
\end{equation}
and 
\begin{equation}
\label{eq:bt}
{\bf B}_T = \nabla \xi \times \nabla \Psi
= \nabla \times \xi \nabla \Psi  = - \nabla \times \Psi \nabla \xi \, .
\end{equation}
The scalar fields $\Phi$ and $\Psi$ are called
the poloidal and toroidal scalar functions, respectively \cite{Backus86}, or Chandrasekhar-Kendall functions 
\cite{Montgomery78, Yoshida91, Low06}. Here $\xi$ is a certain scalar field, which is related to the domain shape. 
As seen in Eq. (\ref{eq:bt}), each field line of the toroidal field ${\bf B}_T$ lies in a constant $\xi$ 
surface (green lines in Fig. 1). On the other hand, Eq. (\ref{eq:bp}) tells that the poloidal field is 
the curl of another toroidal field ${\bf Q}_T  = \nabla \xi \times \nabla \Phi$ and each field line of the poloidal field 
threads through a stack of isosurfaces of $\xi$ (red lines in Fig. 1). 
In astrophysical or geophysical applications, a constant $\xi$ surface usually represents 
a stellar surface or an equipotential surface in a gravitational field. 
In this paper, the 
isosurfaces of the scalar field $\xi$, in each of which the toroidal field line lies, 
will be called the ``toroidal field surfaces.''

\begin{figure}[ht]
\centering
\includegraphics[scale=1.2]{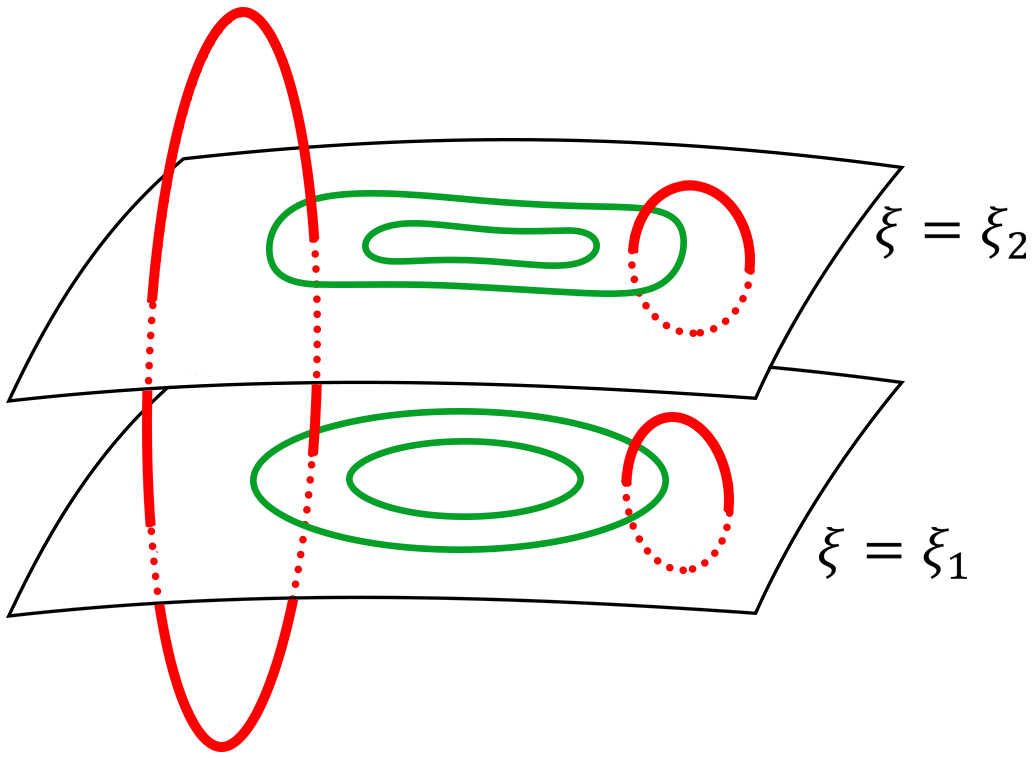}
\caption{Poloidal (red) and toroidal (green) field. Each field line of the toroidal field lies in a constant-$\xi$ surface, which is referred to as the toroidal field surface of 
the poloidal-toroidal representation. Field lines of the poloidal field thread a stack of isosurfaces of $\xi$. In a standard PT representation, the curl of a toroidal field is a poloidal field 
and the curl of a poloidal field should be a toroidal field.}
\label{fig1}
\end{figure}

If the toroidal field surfaces 
are spheres ($\xi=r$ in spherical coordinates) or planes ($\xi=z$ in Cartesian or cylindrical coordinates), 
it can be shown that the curl of 
a poloidal field is again another toroidal field of the form of Eq.~(\ref{eq:bt}) 
\cite{Backus86, Berger18}. The property that the curl of 
a poloidal field is a toroidal field as well as that the curl of a toroidal field is a poloidal field 
is very useful in astrophysical and geophysical applications  
\cite{Elsasser56, Moffatt78, Krause80} and is often considered to be a requirement of a PT representation. 
For an arbitrary scalar field $\xi$, however, the curl of a poloidal field is not necessarily a toroidal field. 
Such a PT representation without any restriction on $\xi$ was named a generalized PT representation \cite{Berger18}. In contrast to this, 
a PT representation, in which the curl of a poloidal field is a toroidal field, will be called a ``standard PT representation.''
Also the toroidal field surfaces of a standard PT representation will be named ``standard toroidal field surfaces.''
Whether a PT representation is standard or generalized depends on the scalar field $\xi$, whose isosurfaces are 
toroidal field surfaces.  
Although it has long been known that spheres and planes are standard toroidal field surfaces of a standard PT representation, the question
whether there are other types of standard toroidal field surfaces has not yet been 
thoroughly addressed. 
This paper is purposed to find a necessary and sufficient condition for isosurfaces of a scalar field $\xi$ to be 
standard toroidal field surfaces so that the curl of a poloidal field of the form of Eq.~(\ref{eq:bp}) 
may be a toroidal field of the form of Eq.~(\ref{eq:bt}), whose field lines lie in these surfaces.

This paper is organized as follows. In the next section, 
a sufficient condition on a scalar field $\xi$ is derived that the curl of a poloidal field should be a toroidal field, i.e., isosurfaces of $\xi$ should be the 
toroidal field surfaces 
of a standard PT representation. In the following section, a necessary condition for it is derived, which is later shown to be identical with the sufficient condition. 
In the succeeding section, we look into the geometrical meaning of 
this necessary and sufficient condition and prove that no standard toroidal field surfaces exist other than 
spheres and planes. 
Then, a discussion on the cylindrical coordinate system is given, and a summary follows to conclude the paper.

\section*{A sufficient condition for the curl of a poloidal field to be a toroidal field}

Let us now consider a three-dimensional (3D) domain, a part of whose boundary is a hypothetical stellar surface or 
an equipotential surface in a gravitational field, and 
we set up a coordinate system there
in such a way that the stellar boundary is a coordinate surface of one coordinate, say, $\xi$.
In this domain, magnetic field is to be described by a PT representation. 
The poloidal field in Eq. (\ref{eq:bp}) and the toroidal field in Eq. (\ref{eq:bt}) may be written in slightly different-looking forms as follows: 
\begin{equation}
\label{eq:bpe}
{{\bf B}_P} 
= \nabla \times \left[ \sum_i { p_i (\xi) \nabla \xi \times \nabla \Phi_i } \right] \, ,
\end{equation} 
and 
\begin{equation}
\label{eq:bte}
{{\bf B}_T} = \sum_j { q_j (\xi) \nabla \xi \times \nabla \Psi_j } \, .
\end{equation}
in which $p_i(\xi)$'s and $q_j(\xi)$'s are arbitrary functions of $\xi$. 
By setting $\displaystyle \Phi = \sum_i p_i (\xi) \Phi_i $ and 
$\displaystyle \Psi = \sum_j q_j (\xi) \Psi_j $, 
Eqs.~(\ref{eq:bpe}) and (\ref{eq:bte}) recover the forms of Eqs.~(\ref{eq:bp}) and (\ref{eq:bt}), respectively. 
Since we are looking for the condition on $\xi$ for a standard PT representation, 
we set 
\begin{equation}
\label{eq:kcp}
{\bf K} = \nabla \times {\bf B}_P = \nabla \times \nabla \times ( \nabla \xi \times \nabla \Phi ) \, ,
\end{equation} 
and seek the condition for ${\bf K}$ to have the form of ${{\bf B}_T}$ in Eq.~(\ref{eq:bte}).
This condition is equivalent to the condition for ${\bf B}_P$ to be of the following form: 
\begin{align}
\label{eq:bpform}
\begin{split}
{\bf B}_P & = \sum_k \eta_k (\xi) \nabla \chi_k + \omega \nabla \xi + \nabla \sigma  \\
& = { \mathrm (A) \ + \ (B) \ + \ (C) } \ , 
\end{split}
\end{align}
in which $\eta_k (\xi)$ is a function of $\xi$, and $\chi_k$, $\omega$ and $\sigma$ are 
arbitrary scalar fields in the domain. Here (A), (B) and (C) respectively stand for the form of each term 
in the right-hand side of Eq.~(\ref{eq:bpform}).
With this form of ${\bf B}_P$, 
it will follow that 
\begin{equation}
\label{eq:kdbp}
{\bf K} = \nabla \times {\bf B}_P = \sum_k { {d \eta_k(\xi) } \over  {d \xi } } \nabla \xi \times \nabla \chi_k
-\nabla \xi \times \nabla \omega \, ,
\end{equation}
whose form is not different from (\ref{eq:bte}). 

Now we examine ${\bf B}_P$ to see if it is of the form (\ref{eq:bpform}). 
\begin{eqnarray}
\label{eq:ccxi}
&{\bf B}_P & = \nabla \times ( \nabla \xi \times \nabla \Phi ) = \nabla \times \nabla \times (\xi \nabla \Phi ) \nonumber \\
&& = \nabla \left[ \nabla \cdot (\xi \nabla \Phi) \right] - \nabla^2 ( \xi \nabla \Phi) \, .
\end{eqnarray}
The first term in the last line of the equation above is of the form (C), which 
contributes nothing when a curl is taken of it. 
The remaining vector Laplacian term can be expanded as 
\begin{equation}
\label{eq:vlapl}
\nabla^2 ( \xi \nabla \Phi) = \xi \nabla (\nabla^2 \Phi) + 2 ( \nabla \xi ) \cdot \nabla \nabla \Phi + (\nabla^2 \xi) \nabla \Phi 
\, .
\end{equation}
The first term in the right-hand side is already of the form (A). 
To handle 
the other terms, 
we introduce an orthogonal coordinate system $(q^1, q^2, q^3)$, in which $q^1 = \xi$. Depending on the shape 
of the $\xi=const.$ surfaces, it may be impossible to set up an orthogonal coordinate system in the whole domain, 
but it is possible at least in the neighborhood of the $\xi=const.$ surface of our interest, e.g., near the stellar boundary. 
Then we have two bases reciprocal (dual) to each other: 
\begin{subequations}
\label{eq:bases}
\begin{align}
\{ {\bf e}^i | {\bf e}^i & = { \nabla q^i}, i=1, 2, 3 \} \, , \\
\{ {\bf e}_i | {\bf e}_i & = { { \partial {\bf r} } \over {\partial q^i} }, i=1, 2, 3 \} \, ,
\end{align}
\end{subequations}
and the components of the metric tensor 
$ g^{ij} = {\bf e}^i \cdot {\bf e}^j $, 
$ g_{ij}  = {\bf e}_i \cdot {\bf e}_j $, and 
$ g_i^j  = {\bf e}_i \cdot {\bf e}^j = \delta_i^j $
are nonzero for $i=j$ only. 
The orthonormal basis $\{ {\hat{\bf e}}_i \}$ is then given by 
\begin{equation}
\label{eq:orthon}
{\hat{\bf e}}_i = {1 \over \sqrt{g_{ii}} } {\bf e}_i 
= {1 \over \sqrt{g^{ii}} } {\bf e}^i \, .
\end{equation}
From now on, we will use the Einstein summation convention, but we will explicitly use summation signs 
when a diagonal component of the metric tensor ($g^{ii}$ or $g_{ii}$) is involved in a summation.  
Since 
$\displaystyle
\nabla = {\bf e}^l { {\partial } \over {\partial q^l } } $
and
$\displaystyle {\bf e}^1 = \nabla q^1 = \nabla \xi$, 
half the second term in the right-hand side of Eq.~(\ref{eq:vlapl}) is expanded as
\begin{align}
(\nabla \xi) \cdot \nabla \nabla \Phi 
& = {\bf e}^1 \cdot {\bf e}^j { {\partial } \over {\partial q^j } }
\left( {\bf e}^i { {\partial \Phi } \over {\partial q^i } } \right)
= g^{11} { {\partial } \over {\partial q^1 } }
\left( {\bf e}^i { {\partial \Phi } \over {\partial q^i } } \right) \nonumber \\
& = g^{11} \left( { {\partial {\bf e}^i } \over {\partial q^1 } } \right)
\left( { {\partial \Phi } \over {\partial q^i } } \right)
+ g^{11} {\bf e}^i { { \partial } \over {\partial q^1 } } 
\left( { {\partial \Phi } \over {\partial q^i } } \right) \, . \label{eq:delxi1}
\end{align}
The last term above is 
$\displaystyle
g^{11} \nabla \left( { {\partial \Phi } \over {\partial q^1 } } \right) $, which 
can be put in the form (A)
if $g^{11}$ is a function of $q^1$ only. 
The condition that 
\begin{equation}
\label{eq:cond1}
g^{11}= \eta (q^1) \ \Leftrightarrow \ |\nabla \xi |^2 = \eta (\xi) \, , 
\tag{Condition I}
\end{equation}
where $\eta$ is any function of $\xi$ only, is named Condition~I. 
Note that Condition~I in the latter expression is free from the 
choice of the coordinate system.

With Condition~I assumed, the first term in the rightmost hand side of Eq.~(\ref{eq:delxi1}) is expanded as
\begin{align}
& g^{11} \left( { {\partial \Phi } \over {\partial q^i } } \right) \left( { {\partial {\bf e}^i } \over {\partial q^1 } } \right) 
= - g^{11} \left( { {\partial \Phi } \over {\partial q^i } } \right) \Gamma_{1k}^i {\bf e}^k
\nonumber \\
= & -{1 \over 2} g^{11} \left( { {\partial \Phi } \over {\partial q^i } } \right) 
g^{im} \left( { { \partial g_{km} } \over {\partial q^1 } } + { { \partial g_{1m} } \over {\partial q^k } } 
- { { \partial g_{1k} } \over {\partial q^m } } \right) {\bf e}^k 
= -{1 \over 2} g^{11} \sum_{i=1}^3 \sum_{k=1}^3 \left[ \left( { {\partial \Phi } \over {\partial q^i } } \right) 
g^{ii} \left( { { \partial g_{ki} } \over {\partial q^1 } } + { { \partial g_{1i} } \over {\partial q^k } } 
- { { \partial g_{1k} } \over {\partial q^i } } \right) {\bf e}^k \right] 
\nonumber \\
= & -{1 \over 2} g^{11} \left[ 
{ \sum_{i=1}^3 \left( { {\partial \Phi } \over {\partial q^i } } \right) 
g^{ii} { { \partial g_{ii} } \over {\partial q^1 } } {\bf e}^i }
+ {\sum_{k=1}^3 \left( { {\partial \Phi } \over {\partial q^1} } \right) 
g^{11} { { \partial g_{11} } \over {\partial q^k } } {\bf e}^k } 
- {\sum_{i=1}^3 \left( { {\partial \Phi } \over {\partial q^i} } \right) 
g^{ii} { { \partial g_{11} } \over {\partial q^i } } {\bf e}^1 }
\right]
\nonumber \\
= \ & - {1 \over 2} g^{11} \sum_{i=1}^3
{ g^{ii} } { { \partial g_{ii} } \over {\partial q^1 } } { {\partial \Phi } \over {\partial q^i } } {\bf e}^i
= {1 \over 2} g^{11} \sum_{i=1}^3
{1 \over g^{ii} } { { \partial g^{ii} } \over {\partial q^1 } } { {\partial \Phi } \over {\partial q^i } } {\bf e}^i
= {1 \over 2} g^{11} \sum_{i=1}^3
{ { \partial \ln g^{ii} } \over {\partial q^1 } } { {\partial \Phi } \over {\partial q^i } } {\bf e}^i \nonumber \\
= \ &
{1 \over 2} g^{11} { { \partial \ln g^{11} } \over {\partial q^1 } } { {\partial \Phi } \over {\partial q^1 } } \nabla q^1
+ {1 \over 2} g^{11} { { \partial \ln g^{22} } \over {\partial q^1 } } { {\partial \Phi } \over {\partial q^2 } } \nabla q^2
+ {1 \over 2} g^{11} { { \partial \ln g^{33} } \over {\partial q^1 } } { {\partial \Phi } \over {\partial q^3 } } \nabla q^3 \, ,
\label{eq:longex}
\end{align}
in which $\Gamma_{1k}^i$ is a Christoffel symbol of the second kind. In the above development, we have exploited 
Condition I that 
\begin{equation*}
{ { \partial g^{11 } } \over {\partial q^2 } } = { { \partial g^{11 } } \over {\partial q^3 } } = 0
\end{equation*}
as well as
the properties of the metric tensor in orthogonal coordinate systems such as 
\begin{equation*}
- g^{ii} { { \partial g_{ii} } \over {\partial q^1 } }
= {1 \over g^{ii} } { { \partial g^{ii} } \over {\partial q^1 } }
= { { \partial \ln g^{ii} } \over {\partial q^1 } } \, .
\end{equation*}
At a glance of the last line of Eq.~(\ref{eq:longex}), one may notice that if 
\begin{equation}
\label{eq:cond2o}
{ { \partial \ln g^{22} } \over {\partial q^1 } }
= { { \partial \ln g^{33} } \over {\partial q^1 } } = {f} (q^1)
= { { \partial \ln g^{11} } \over {\partial q^1 } } \, ,
\end{equation}
in which $f$ is a function of one independent variable,
it could be put in the form (A). 
That condition is indeed a sufficient condition for it to take the form (A), but is too restrictive to accept hastily. 
It should be noted that the first term in the last line of Eq.~(\ref{eq:longex}) is already in 
the form (B) since $q^1=\xi$. If 
\begin{equation}
\label{eq:cond2}
{ { \partial \ln g^{22} } \over {\partial q^1 } }
= { { \partial \ln g^{33} } \over {\partial q^1 } } = {f} (q^1) \, ,
\end{equation}
where the last equality in Eq.~(\ref{eq:cond2o}) has been abandoned, then
the rightmost hand side of Eq.~(\ref{eq:longex}) can be rewritten as
\begin{align} 
& {1 \over 2} g^{11} (q^1) { { \partial \ln g^{11} } \over {\partial q^1 } } { {\partial \Phi } \over {\partial q^1 } } \nabla q^1
+ {1 \over 2} g^{11} (q^1) {f} (q^1) \left[ { {\partial \Phi } \over {\partial q^2 } } \nabla q^2
+ { {\partial \Phi } \over {\partial q^3 } } \nabla q^3 \right]
\nonumber \\
= \ & {1 \over 2} g^{11} \left[ { { \partial \ln g^{11} } \over {\partial q^1 } }  
- {f} (q^1) \right] { {\partial \Phi } \over {\partial q^1 } } \nabla q^1
+ {1 \over 2} g^{11} {f} (q^1) \sum_{i=1}^3 { {\partial \Phi } \over {\partial q^i } } \nabla q^i \, .
\label{eq:longex2}
\end{align}
In the right-hand side, the first term with $\nabla q^1$ is of the form (B) and the second term with a summation is of the form (A). 
We name the condition given by Eq.~(\ref{eq:cond2}) Condition II.
One can see the following equivalence 
\begin{align}
\label{eq:cond2a}
\tag{Condition II}
\begin{split}
& { { \partial \ln g^{22} } \over {\partial q^1 } }
= { { \partial \ln g^{33} } \over {\partial q^1 } } = {f} (q^1) \\
\Leftrightarrow \ \ 
& g^{22} (q^1, q^2, q^3) = {\mathcal F} (q^1) {\mathcal G} (q^2, q^3) \quad {\mathrm {and}} \\
& g^{33} (q^1, q^2, q^3) = {\mathcal F} (q^1) {\mathcal H} (q^2, q^3) \, ,
\end{split}
\end{align}
in which $ {\mathcal F} (q^1)$ is a function of one independent variable $q^1$, 
and ${\mathcal G} (q^2, q^3)$ and ${\mathcal H} (q^2, q^3)$ are functions of two independent variables $q^2$ and $q^3$. 
Thus, $g^{22}$ and $g^{33}$ must respectively be factorized into a $q^1$-dependent part and a $(q^2, q^3)$-dependent 
part, and $g^{22}$ and $g^{33}$ must share the same $q^1$-dependent factor. 
Now we only need to address the last term in the right-hand side of Eq.~(\ref{eq:vlapl}). 
The term can be put in the form (A) if 
\begin{equation}
\label{eq:laplxi}
\nabla^2 \xi = {\tilde \eta} (\xi) \, ,
\end{equation}
in which ${\tilde \eta} (\xi)$ is any function of one independent variable $\xi$. 
At this point, we are to raise the question whether the condition of Eq.~(\ref{eq:laplxi}) 
is independent of Conditions I and II. Let us expand $\nabla^2 \xi$ in the orthogonal coordinate
system as we have set up above. 
\begin{align}
& \nabla \cdot \nabla \xi 
= {\bf e}^j { {\partial } \over {\partial q^j } } \cdot {\bf e}^k { {\partial q^1 } \over {\partial q^k } }
= {\bf e}^j \cdot { {\partial {\bf e}^1 } \over {\partial q^j } } = -\sum_j g^{jj} \Gamma_{jj}^1 \nonumber \\
= \ & {1 \over 2} g^{11} \left[
{ { \partial \ln g^{11} } \over {\partial q^1 } }
- { { \partial \ln g^{22} } \over {\partial q^1 } }
- { { \partial \ln g^{33} } \over {\partial q^1 } } \right]
= {1 \over 2} g^{11} 
{ { \partial } \over {\partial q^1 } } 
\left( \ln { { g^{11} } \over { {g^{22} } {g^{33} } } } \right) \, .
\label{eq:laplxi2}
\end{align}
The condition for the last expression to be a function of $q^1$ only is that 
$g^{11}$, $g^{22}$ and $g^{33}$ are respectively factorized into a $q^1$-dependent function and 
a $(q^2, q^3)$-dependent function, which is satisfied if Conditions I and II are both met. Thus, Conditions I and II 
combined are a sufficient condition for Eq.~(\ref{eq:laplxi}), but might not be a necessary condition because the former 
specify more details than the latter. From the above analysis, we can conclude that if Conditions I and II are both met, 
the curl of a poloidal field takes the form of a toroidal field as given by Eq.~(\ref{eq:bte}). 
Thus, Conditions I and II combined are a sufficient condition for the curl of 
a poloidal field to be a toroidal field.

\section*{A necessary and sufficient condition for the curl of a poloidal field to be a toroidal field}

It is still uncertain whether Conditions I and II combined are also a necessary condition for the curl of 
a poloidal field to be a toroidal field. 
In order to check this, we will seek the condition for 
\begin{equation}
\label{eq:curlbp}
\nabla \xi \cdot \nabla \times {\bf B}_P = 0 \, . 
\end{equation}
If $\nabla \times {\bf B}_P$ is a toroidal field given by Eq.~(\ref{eq:bte}), 
Eq.~(\ref{eq:curlbp}) surely holds, but it is not transparent whether Eq.~(\ref{eq:curlbp})
guarantees that $\nabla \times {\bf B}_P$ is a toroidal field having the form of Eq.~(\ref{eq:bt}) or (\ref{eq:bte}). 
Thus, we can safely say that Eq.~(\ref{eq:curlbp}) is 
a necessary condition for $\nabla \times {\bf B}_P$ to be a toroidal field   
while Conditions I and II combined are a sufficient condition for it. 
Here we want to find a condition equivalent to Eq.~(\ref{eq:curlbp}) and compare it with 
Conditions I and II. 
In an orthogonal coordinate system, we will directly calculate 
\begin{equation*}
{\bf e}^1 \cdot \nabla \times {\bf B}_P = {\bf e}^1 \cdot \nabla \times \nabla \times {\bf Q}_T
\end{equation*}
to seek the condition for this expression to be zero. 
Here
\begin{align}
{\bf Q}_T & = \nabla q^1 \times \nabla \Phi = {\bf e}^1 \times {\bf e}^i { {\partial \Phi } \over {\partial q^i } } \nonumber \\
& = { \sqrt{g^*} \over g^{33} } { {\partial \Phi } \over {\partial q^2 } } {\bf e}^3 
- { \sqrt{g^*} \over g^{22} } { {\partial \Phi } \over {\partial q^3 } } {\bf e}^2
= \sqrt{ {g_{33}} \over {g } } { {\partial \Phi } \over {\partial q^2 } } {\hat {\bf e}}_3 
- \sqrt{ {g_{22}} \over {g } } { {\partial \Phi } \over {\partial q^3 } } {\hat {\bf e}}_2 \, ,
\label{eq:qtcomp}
\end{align}
in which 
\begin{equation}
\label{gnumber}
g^* = g^{11} g^{22} g^{33} = {1 \over {g_{11} g_{22} g_{33} } } = {1 \over g} \, .
\end{equation}
After some tedious algebra, we have 
\begin{align}
& - \sqrt{1 \over g^*} \,  {\bf e}^1 \cdot \nabla \times \nabla \times {\bf Q}_T \nonumber \\
= & { {\partial } \over {\partial q^2 } } \left[
{ \sqrt{g^*} \over g^{33} }{ {\partial } \over {\partial q^1 } } \left(
{ \sqrt{g^*} \over g^{22} }{ {\partial \Phi } \over {\partial q^3 } }
\right)
\right]
- { {\partial } \over {\partial q^3 } } \left[
{ \sqrt{g^*} \over g^{22} }{ {\partial } \over {\partial q^1 } } \left(
{ \sqrt{g^*} \over g^{33} }{ {\partial \Phi } \over {\partial q^2 } }
\right)
\right] 
\nonumber \\
= & \left[ { {\partial g^{11} } \over {\partial q^2 } } \right]
{ {\partial^2 \Phi } \over {\partial q^3 \partial q^1} }
- \left[ { {\partial g^{11} } \over {\partial q^3 } } \right]
{ {\partial^2 \Phi } \over {\partial q^1 \partial q^2} }
\nonumber \\
+ & \left[ 
{ \sqrt{g^*} \over g^{33} }{ {\partial } \over {\partial q^1 } } 
\left( { \sqrt{g^*} \over g^{22} } \right)
- { \sqrt{g^*} \over g^{22} }{ {\partial } \over {\partial q^1 } } 
\left( { \sqrt{g^*} \over g^{33} } \right)
\right] { {\partial^2 \Phi } \over {\partial q^2 \partial q^3} }
\nonumber \\
+ & \left[ { {\partial } \over {\partial q^2 } }
\left( { \sqrt{g^*} \over g^{33} }{ {\partial } \over {\partial q^1 } } 
\left( { \sqrt{g^*} \over g^{22} } \right) \right)
\right]{ {\partial \Phi } \over {\partial q^3 } }
- \left[ { {\partial } \over {\partial q^3 } }
\left( { \sqrt{g^*} \over g^{22} }{ {\partial } \over {\partial q^1 } } 
\left( { \sqrt{g^*} \over g^{33} } \right) \right)
\right]{ {\partial \Phi } \over {\partial q^2 } } \, .
\label{eq:ccqt}
\end{align}
For this expression to be identically zero for an arbitrary $\Phi$, the five coefficients 
in square brackets in the rightmost hand side of the equation must be all zero. The first two coefficients being zero implies that 
$g^{11}$ must be a function of $q^1$ only, which is nothing but our Condition I. 
The third coefficient term can be rewritten as
\begin{align}
& { \sqrt{g^*} \over g^{33} }{ {\partial } \over {\partial q^1 } } 
\left( { \sqrt{g^*} \over g^{22} } \right)
- { \sqrt{g^*} \over g^{22} }{ {\partial } \over {\partial q^1 } } 
\left( { \sqrt{g^*} \over g^{33} } \right)
\nonumber \\
= & {g^* \over {g^{22} g^{33} }} { {\partial } \over {\partial q^1 } } 
\ln { { \sqrt{g^*} / g^{22} } \over { { \sqrt{g^*} / g^{33} } }}
= g^{11} { {\partial } \over {\partial q^1 } } \ln { { g^{33} } \over { g^{22} } } \, .
\label{eq:thirdc}
\end{align}
The condition for this to be zero is the same as Condition II. 
Under Conditions I and II, i.e., under the condition that the first three coefficients of Eq.~(\ref{eq:ccqt}) be zero, 
the fourth coefficient term becomes
\begin{align}
& { {\partial } \over {\partial q^2 } }
\left( { \sqrt{g^*} \over g^{33} }{ {\partial } \over {\partial q^1 } } 
\left( { \sqrt{g^*} \over g^{22} } \right) \right)
= { {\partial } \over {\partial q^2 } }
\left( g^{11} { {\partial } \over {\partial q^1 } } 
\ln { \sqrt{g^*} \over g^{22} } \right)
= {1 \over 2} { {\partial } \over {\partial q^2 } }
\left( g^{11} { {\partial } \over {\partial q^1 } } 
\ln { { g^{11} g^{33} } \over g^{22} } \right)
\nonumber \\
= & {1 \over 2} { {\partial } \over {\partial q^2 } }
\left( { {\partial g^{11} } \over {\partial q^1 } } 
\right)
= 0 \, .
\label{eq:fourthc}
\end{align}
In the same way, the fifth coefficient term of Eq.~(\ref{eq:ccqt}) is zero. 
Therefore, Conditions I and II combined are equivalent to the condition for 
Eq.~(\ref{eq:curlbp}) to hold, which is a necessary condition for 
$\nabla \times {\bf B}_P$ to be a toroidal field. Since we have already seen that Conditions I and II combined 
are a sufficient condition for it, they are the necessary and sufficient condition for the curl of a poloidal field to be 
a toroidal field.

\section*{Geometrical meaning of the condition derived above}

\begin{figure}[ht]
\centering
\includegraphics[scale=0.55]{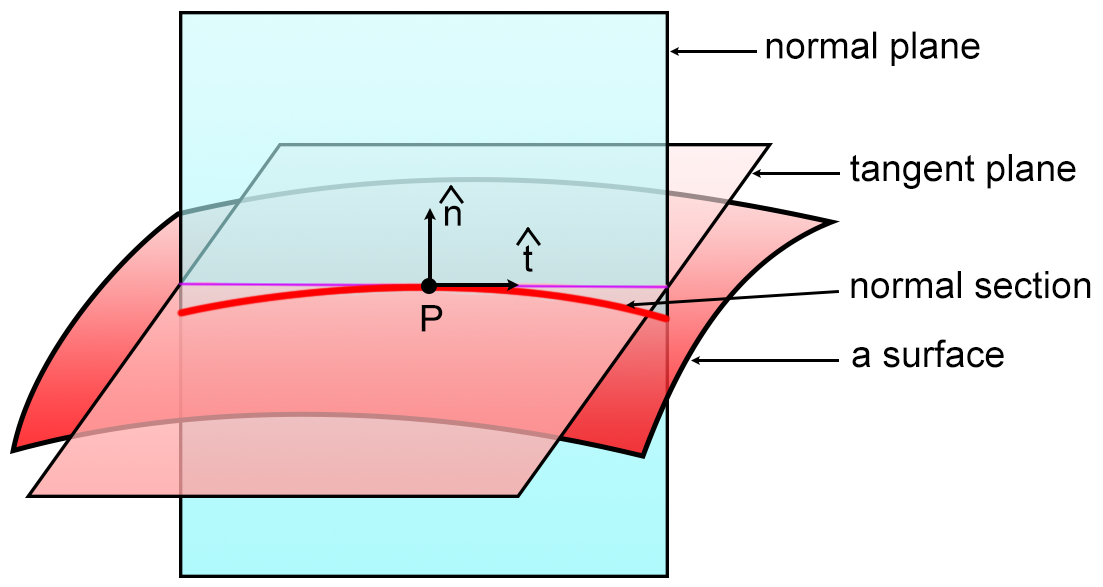}
\caption{Normal section of a surface. A normal plane is spanned by a normal vector $\hat {\bf n}$ and a tangent vector $\hat {\bf t}$ to the surface at a point $P$ in the surface. 
The intersection of 
the surface and the normal plane is a normal section. There are infinitely many normal sections passing through the point $P$. 
The curvature of a normal section is a normal curvature, which is a function of $P$ and $\hat {\bf t}$. }
\label{fig2}
\end{figure}

What is then the geometrical meaning of Conditions I and II? The coordinate-free expression of Condition I, 
$|\nabla \xi|^2=\eta(\xi)$, tells that all constant-$\xi$ surfaces are parallel surfaces \cite{Gray06}. 
One can draw parallel surfaces in the neighborhood of 
any continuous surface. The condition does not mean similarity of constant-$\xi$ surfaces. For example, 
parallel planes, co-axial cylinders and concentric spheres are respectively similar and parallel to each other, but 
confocal ellipsoids, though similar, are not parallel to each other while 
parallel surfaces of an ellipsoid are not similar to each other. 
In contrast to Condition I, Condition II is apparently given in a coordinate language, 
but we want to translate it into a geometrical (coordinate-free) 
language. For the time being, we will hold to an orthogonal coordinate system with $q^1=\xi$. 
Then, a unit normal vector to a $\xi=const.$ surface is 
\begin{equation}
\label{eq:nvec}
{\hat {\bf n}} = {\hat {\bf e}}_1= 1 / \sqrt{g^{11} } {\bf e}^1 \, ,
\end{equation}
and an arbitrary unit tangent vector to the surface is represented by
\begin{equation}
\label{eq:tvec}
{\hat {\bf t}} = \sum_{j=2}^3 t_j {\bf e}^j = \sum_{k=2}^3 t^k {\bf e}_k \, .
\end{equation}
Since ${\hat {\bf t}}$ is a unit vector,
\begin{equation}
\label{eq:tunit}
{\hat {\bf t}} \cdot {\hat {\bf t}} = \sum_{j=2}^3 t_j t^j = 1 \, .
\end{equation}
The normal vector ${\hat {\bf n}}$ and a tangent vector ${\hat {\bf t}}$ to a constant-$\xi$ surface span 
a so-called normal plane (see Fig.~2). The intersection of the surface and a normal plane is a curve called normal section. 
The curvature $\kappa_n$ of a normal section is a normal curvature \cite{Sochi17}, which is given by 
\begin{equation}
\label{eq:kappan}
\kappa_n ({\bf r}, {\hat {\bf t}}) = - {\hat {\bf t}} \cdot { { d {\hat {\bf n}} } \over ds } 
= - {\hat {\bf t}} \cdot \left( {\hat {\bf t}} \cdot \nabla {\hat {\bf n}} \right) \, ,
\end{equation}
in which $ds$ is the arclength element of the normal section in the ${\hat {\bf t}}$-direction and
${ { d {\hat {\bf n}} } / ds } = {\hat {\bf t}} \cdot \nabla {\hat {\bf n}} $ is the directional 
derivative of ${\hat {\bf n}}$ in that direction. 
To find $\kappa_n$ in a constant-$\xi$ surface, we use the following calculations.
Under Condition I, we have 
\begin{equation}
\label{eq:tdotg}
{\hat {\bf t}} \cdot \nabla {f} \left( g^{11} (q^1) \right) = 0, 
\end{equation}
in which $ {f}$ is an arbitrary function of one independent variable. 
Under Conditions I and II both, we have for $j=2, 3$, 
\begin{align}
& {\bf e}^j \cdot \nabla {\bf e}^1 
= g^{jj} { { \partial {\bf e}^1 } \over { \partial q^j } } 
= - g^{jj} \sum_{l=1}^3 \Gamma_{jl}^1 {\bf e}^l
\nonumber \\
= & {1 \over 2} g^{11} g^{jj} { { \partial g_{jj} } \over { \partial q^1 } } {\bf e}^j
= - {1 \over 2} g^{11} { { \partial \ln g^{jj} } \over { \partial q^1 } } {\bf e}^j
= - {1 \over 2} g^{11} { { \partial \ln {\mathcal F} (q^1) } \over { \partial q^1 } } {\bf e}^j \, .
\label{eq:ejdele1} 
\end{align}
Using Eqs.~(\ref{eq:nvec})-(\ref{eq:ejdele1}), 
we find
\begin{align}
\kappa_n 
& = - {\hat {\bf t}} \cdot \left[ {\hat {\bf t}} \cdot \nabla \left( { 1 \over \sqrt{g^{11}} } {\hat {\bf e}}_1 \right) \right]
\nonumber \\ 
& = - { 1 \over \sqrt{g^{11}} } \sum_{k=2}^3 \sum_{j=2}^3 t^k {\bf e}_k \cdot \left(
t_j {\bf e}^j \cdot \nabla {\bf e}^1 \right)
= {1 \over 2} \sqrt{g^{11}} \sum_{k=2}^3 \sum_{j=2}^3 t^k t_j {\bf e}_k \cdot 
\left( { { \partial \ln {\mathcal F} (q^1) } \over { \partial q^1 } } {\bf e}^j \right)
\nonumber \\ 
& = {1 \over 2} \sqrt{g^{11}} { { \partial \ln {\mathcal F} (q^1) } \over { \partial q^1 } }
\sum_{j=2}^3 t^j t_j
= {1 \over 2} \sqrt{g^{11}} { { \partial \ln {\mathcal F} (q^1) } \over { \partial q^1 } } \, ,
\label{eq:kapn}
\end{align}
which is a function of $q^1 = \xi$ only and does not depend on the position in the surface nor on the direction 
of the normal section. 
Therefore, Conditions I and II geometrically imply that the normal curvatures in all directions  
at all points in a constant-$\xi$ surface should be the same. Among all 2D surfaces embedded in a 3D Euclidean space, 
only spheres and planes have this property. 
Thus, a standard PT representation, which is formulated by either Eqs.~(\ref{eq:bp})-(\ref{eq:bt}) or (\ref{eq:bpe})-(\ref{eq:bte})
and in which the curl of a poloidal field is a toroidal field, 
is possible 
for $\xi = {f}(r)$, where $r$ is the radial distance from 
a certain point (e.g., the center of a star) and ${f}$ is 
a generic function of one independent variable, or for $\xi={f}(z)$, where $z$ is the normal distance from a plane 
(e.g., a stellar surface approximated by a plane). 
It is thus not surprising that a standard PT representation has so far been employed only in spherical, Cartesian 
or cylindrical coordinate systems.

\section*{Discussion on cylindrical coordinate systems}

In a cylindrical coordinate system 
$(q^1, q^2, q^3) = (\rho, \varphi, z)$, $(\varphi, z, \rho)$ or $(z, \rho, \varphi)$, 
we have 
$g^{\rho \rho} =1$, $g^{\varphi \varphi}= \rho^{-2}$ and $g^{z z}=1$. 
The choice $q^1=z$ satisfies Conditions~I and II both and the
parallel planes $z=const.$ are qualified for standard toroidal field surfaces. 
The choice $q^1=\varphi$ does not satisfy Condition I, and the isosurfaces of $\varphi$ are not 
parallel surfaces. The choice $q^1=\rho$ satisfies Condition I, but not Condition II because the $\rho$-dependent
factors of $g^{\varphi \varphi}$ and $g^{z z}$ are not identical, which corresponds to the geometrical observation that
the normal curvature at each point of a cylindrical surface is zero in the axial direction, 
but nonzero and varying in other directions. Therefore, the co-axial cylindrical surfaces cannot be  
standard toroidal field surfaces. 

Here one may be puzzled at the last statement, seeing that the terms ``poloidal'' and ``toroidal'' are most commonly
used referring to cylindrical or toroidal laboratory plasmas. 
If one considers a magnetic field with flux surfaces of a torus shape, whose axis of revolution is the $z$-axis, 
then the toroidal field lies in $z=const.$ planes and the poloidal field in 
planes of constant azimuth, not different from our sense of those terms. In laboratory plasmas, however, both the toroidal 
field and the poloidal field are expressed in the form of our toroidal field (Eq. [\ref{eq:bt}] or [\ref{eq:bte}]). For example,
\begin{equation}
\label{eq:tplab}
{\bf B}={1\over {2\pi}} { { d \Psi_{tor} } \over {d {\tilde \rho}} } \nabla {\tilde \rho} \times \nabla \theta_f
+ {1\over {2\pi}} { { d \Psi_{pol} } \over {d {\tilde \rho}} } \nabla {\tilde \rho} \times \nabla \zeta_f \, ,
\end{equation}
in which $\Psi_{tor}$ and $\Psi_{pol}$ are respectively the toroidal flux enclosed by, and the poloidal flux outside the flux surface labeled by ${\tilde \rho}$, 
and $\theta_f$ and $\zeta_f$ are respectively generalized poloidal and toroidal angles \cite{Dhaeseleer91}. 
In our definition of the poloidal and toroidal fields 
(Eqs.~[\ref{eq:bp}]-[\ref{eq:bt}] or Eqs.~[\ref{eq:bpe}]-[\ref{eq:bte}]), neither $\xi$ nor $\Phi$ nor $\Psi$ needs to be a flux surface label for the total ${\bf B}$.  
If we narrow down the definition of the poloidal and toroidal fields to such that the curl of a toroidal field is a poloidal field and 
the curl of a poloidal field a toroidal field, each term in equation~(\ref{eq:tplab}) is qualified for a toroidal or poloidal field,  
only if a magnetic flux surface is also a current surface, i.e., 
${\bf B}\cdot \nabla {\tilde \rho} = 0$ and ${\bf J}\cdot \nabla {\tilde \rho} = 0$, 
which is possible only in a magnetohydrodynamic (MHD) equilibrium ${\bf J}\times {\bf B} - \nabla p =0$.
Therefore, the label of a cylindrical surface or a toroidal surface can be our scalar field $\xi$ in Eqs.~(\ref{eq:bp})-(\ref{eq:bt}) only under
very special conditions, which cannot be generally applied for all magnetic fields. 

\section*{Summary}

In this paper, we have derived a necessary and sufficient condition on the scalar field $\xi$ in the standard poloidal-toroidal representation (Eqs.~[\ref{eq:bpt}]-[\ref{eq:bt}])
that the curl of a poloidal field should be a toroidal field. It is given by Conditions I and II combined. Its geometrical meaning is that each isosurface of $\xi$ must have a constant normal curvature in all directions at all points. In a 3D Euclidean space, only spheres and planes satisfy this condition. 
Thus, there can be no toroidal field surfaces for the standard PT representation other than spheres and planes. 
The poloidal-toroidal conversion through a curl operation, therefore, can be done only in an approximate sense if a PT representation is used for describing 
dynamos or other magnetic processes in a celestial body of a highly oblate shape. 
However, exotic surfaces corresponding to our standard toroidal field surfaces might be available in dimensions more than three or in non-Euclidean spaces, e.g., 
in a curved 4D spacetime, 
which is, though intriguing, far beyond the scope of the present study. 

\section*{Methods}

We have used vector and tensor analysis with differential geometry of curves and surfaces.

\bibliography{references}

\section*{Acknowledgements}

This work was supported by the National Research Foundation of Korea (NRF) Grant 2019R1F1A1060887 funded by the Ministry of Science and ICT of the Korean government.

\section*{Author contributions statement}

G.S.C. recognized the importance of the problem addressed in the paper. S.Y. and G.S.C. together performed mathematical calculations. 
The manuscript is cooperatively written by the two authors. 
G.S.C. secured the funding for the research.

\section*{Competing interests}

The authors declare no competing interests.

\section*{Additional information}

\textbf{Correspondence} should be addressed to G.S.C.

\bigskip

\noindent
After the submission of the present paper, it has been brought to our notice that Dr.~J.~J.~Aly (2022, to be submitted) has independently reached the same conclusion as ours. 
His mathematical techniques are different from ours presented in this paper. 

After the publication of the present paper, Dr.~Matthias Rheinhardt kindly directed the authors' attention to a paper \cite{Radler74}, which we the authors had not been aware of. Its Section 6 dealt with the same subject and reached the same conclusion as ours. However, it and our paper, respectively, have taken different mathematical routes to attain the results.

\end{document}